\documentclass[useAMS,usenatbib]{mn2e}
\usepackage{latexsym}
\usepackage{url}
\usepackage{graphicx}
\usepackage{hyperref}

\mathchardef\mhyphen="2D

\newcommand{\bq}{\textbf{q}}

\newcommand{\bx}{\textbf{x}}

\newcommand{\hmpc}{\,$h^{-1}$\,Mpc}

\newcommand{\LCDM}{$\Lambda$CDM}

\newcommand{\eg}{e.g., }

\newcommand{\bPsi}{\mbox{\boldmath $\Psi$}}

\newcommand{\org}{{\scshape origami}}
\newcommand{\nstreams}{{N_{\rm streams}}}

\chardef\til=`\~

\begin{document}

\title{Origami constraints on the initial-conditions arrangement of
  dark-matter caustics and streams}

\author[Mark C.\ Neyrinck]
{Mark C.\ Neyrinck$^1$\\
$^{1}$Department of Physics and Astronomy, The Johns Hopkins University, Baltimore, MD 21218, USA}



\maketitle

\begin{abstract}
In a cold-dark-matter universe, cosmological structure formation
proceeds in rough analogy to origami folding. Dark matter occupies a
three-dimensional `sheet' of free-fall observers, non-intersecting in
six-dimensional velocity-position phase space.  At early times, the
sheet was flat like an origami sheet, i.e.\ velocities were
essentially zero, but as time passes, the sheet folds up to form
cosmic structure.  The present paper further illustrates this analogy,
and clarifies a Lagrangian definition of caustics and streams:
caustics are two-dimensional surfaces in this initial sheet along
which it folds, tessellating Lagrangian space into a set of
three-dimensional regions, i.e.\ streams.  The main scientific result
of the paper is that streams may be colored by only two colors, with
no two neighbouring streams (i.e.\ streams on either side of a caustic
surface) colored the same.  The two colors correspond to positive
and negative parities of local Lagrangian volumes.  This is a severe
restriction on the connectivity and therefore arrangement of streams
in Lagrangian space, since arbitrarily many colors can be necessary
to color a general arrangement of three-dimensional regions.  This
stream two-colorability has consequences from graph theory, which we
explain.  Then, using $N$-body simulations, we test how these caustics
correspond in Lagrangian space to the boundaries of haloes, filaments
and walls.  We also test how well outer caustics correspond to a
Zel'dovich-approximation prediction.
\end{abstract}

\begin {keywords}
  large-scale structure of Universe -- cosmology: theory
\end {keywords}

\section{Introduction}
On moderately large scales, matter and galaxies in the Universe trace
what is known as a cosmic web \citep[\eg][]{BondEtal1996}.  Regions
well under the mean density develop into voids.  At higher density,
the matter is arranged into planar structures known as walls or
pancakes.  Still higher-density matter has a filamentary morphology.
At the highest densities, it is assembled into pointlike clusters, or
haloes.

There are several ways to characterize these structures.
Observationally, one way of identifying and defining them is by
looking for depressions, ridges, and peaks in the (over)density field
$\delta$.  For example, one can measure the eigenvalues of the Hessian
$\partial^2\delta/\partial x_i\partial x_j$; this gives local density
maxima/ridges in one, two and three dimensions
\citep[\eg][]{HahnEtal2006,Miguel2007,SousbieEtal2008,PogosyanEtal2009,Sousbie2011}.
Another approach is global, defining voids to tessellate space.  For
instance, they can be defined as density depressions outlined by a
watershed transform \citep{Platen2007,Neyrinck2008}.  In this
framework, walls, filaments, and haloes are defined according to where
voids meet each other, and the dimensionality of borders separating
them \citep{Miguel2010}. In cosmology, these two definitions turn out
to be rather similar, but this only is true in detail if no locally
defined walls and filaments end within voids.

Another way to understand the structures is as folds in a
three-dimensional manifold (`sheet'), a Lagrangian picture.  In
Eulerian space, densities and velocities at fixed positions are
tracked, but particles move around.  But in Lagrangian space, each
dark-matter parcel retains the same coordinates.  First-order
Lagrangian perturbation theory, known as the \citet{Zeldovich1970}
approximation, already is quite useful to understand the basics of
cosmological structure formation.

In this Lagrangian framework, particles in an $N$-body simulation can
be thought of not simply as blobs of mass, but as vertices on an
initially cubic grid that gravity deforms, and eventually causes to
self-cross in three-dimensional position space.  In six-dimensional
velocity-position phase space, though, the manifold does not cross
itself, a property which prompted our recent analogy to the
paper-folding that occurs in origami \citep{FalckEtal2012}.  In that
paper, we introduced a structure-finding algorithm (not to be confused
with the origami analogy itself) called \org\ (Order-ReversIng
Gravity, Apprehended Mangling Indices) to identify structures in an
$N$-body simulation according to the number of axes along which the
initial-conditions (Lagrangian) lattice has crossed itself.  We
briefly describe this algorithm below in the description of
Fig.\ \ref{fig:morphparity1}.  A couple of other recent papers
\citep{ShandarinEtal2011, AbelEtal2011} have also explored the power
of working within the dark-matter mesh.  \citet{ShandarinEtal2011}
identify stream-crossings with overlapping tetrahedra from a
tessellation of initial Lagrangian space. \citet{AbelEtal2011} explore
the benefits of measuring the density natively within the phase-space
sheet, in principle greatly reducing particle-discreteness problems.

Another major application of a Lagrangian framework is the study of
caustics, largely the subject of present work.  A `caustic' is the
edge of a fold of the sheet in six-dimensional phase space.  A strong
observational motivation for studying caustics is that dark matter, if
it annihilates, may do so most often in them, since formally the
density goes arbitrarily high there.  Thus they may greatly enhance
observable signals from dark matter
\citep[\eg][]{Hogan2001,NatSik2008,VogelsbergerWhite2011}

The behavior of caustics in Eulerian space has been extensively
studied.  Catastrophe theory allows a rigorous description and
classification of the kinds of caustics and structures that can
develop in Eulerian space when the manifold folds and crosses itself
\citep[\eg][]{ArnoldEtal1982,Arnold2001}.  

A full understanding of the dynamics of the phase space sheet requires
consideration of its symplectic structure, a special kind of geometry
in which the position and velocity (sub)spaces do not directly mix on
an equal footing.  For example, the metric in a symplectic space
considers the position and velocity spaces seperately.  In this paper,
we do not explicitly consider the velocity subspace, concentrating on
the spatial displacements between Lagrangian and Eulerian (both
position) space.  We note in passing, though, that the full dynamics
of the sheet has special physical significance in general relativity,
since the sheet consists of the set of observers that have experienced
free-fall throughout cosmic time.

A Lagrangian view of the cosmic web, as we mainly adopt here, is
difficult to apply directly to observed data; however, perhaps methods
that construct initial conditions from observed data could in
principle allow an avenue to do so
\citep[e.g.][]{MohayaeeEtal2006,LavauxEtal2008}.  And, for example,
tracking caustics in simulations is relevant to the caustic method of
estimating galaxy-cluster masses, which uses the outer caustics of
cluster haloes as observed in redshift space
\citep{DiaferioGeller1997}.

This paper is organized as follows.  In Section \ref{sec:origami}, we
briefly review some properties of origami mathematics that are
applicable in cosmology.  In Section \ref{sec:oricosmi}, we apply
these properties to cosmology, giving Lagrangian definitions of the
concepts of `caustics' and `streams.'  We also explain some
graph-theory consequences of the observation that streams are
two-colorable according to their parity, and show some relevant
measurements from $N$-body simulations.

\section{Origami mathematics}
\label{sec:origami}

Considering how ancient the art of origami is, the mathematics of it
have developed relatively recently.  A relatively early result (dating
to 1893) is that paper-folding is a more powerful mechanism for
solving geometric and algebraic problems than the classical ruler and
compass.  For example, in principle it can be used to trisect an
angle, and to solve some types of cubic
equations\citep{Row1966,Martin1998,Hull2006}.  Then, in the last few
decades, significant advances were made in algorithmic origami design
\citep[\eg][]{Lang2003}.

\begin{figure}
  \begin{center}
    \includegraphics[scale=0.43]{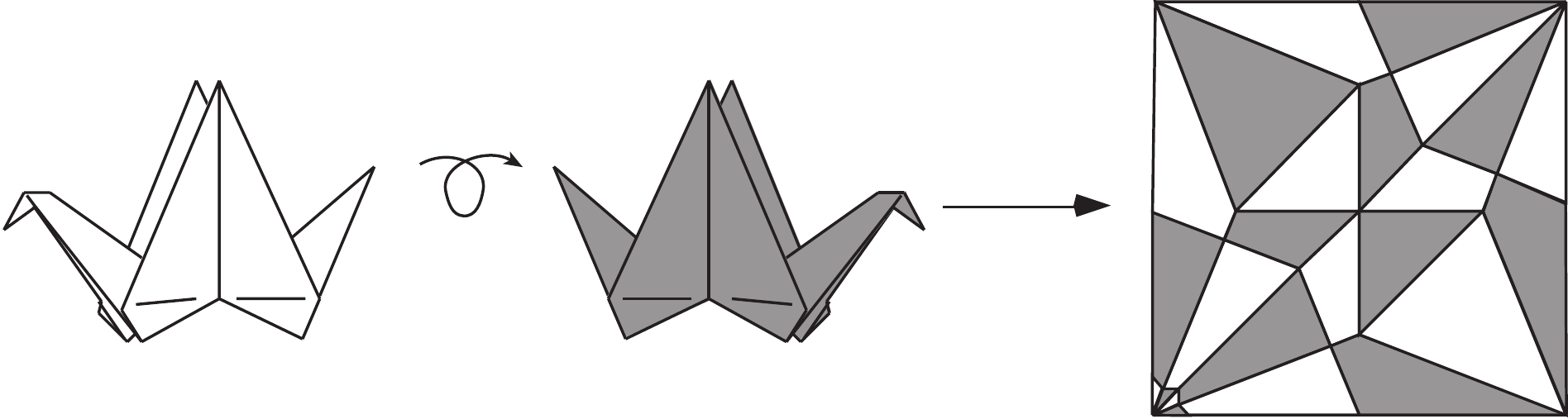}
  \end{center}  
  \caption{Two-coloring of the polygons outlined by creases in an
    origami `traditional Japanese flapping bird' (similar to a
    crane). Polygons facing `up' (out of the page in the leftmost
    diagram, with the head facing to the left) are colored white, while
    polygons facing `down' (into the page in the leftmost diagram) are
    painted gray.  The crane has been unfolded for the rightmost
    diagram.  Creases are shown as lines here; the polygons outlined
    by them are successfully two-colored.  Figure from
    \citet{Hull2006}, courtesy Tom Hull.
    \label{fig:twocolor}
  }
\end{figure}

Here we focus on some results in `flat origami' that are particularly
relevant to large-scale structure.  In flat origami, folding of a
two-dimensional sheet is allowed in three dimensions, but the result
is restricted to lie flat in a plane, i.e.\ it could be squashed
between pages in a book without acquiring any new creases.  The class
of flat-foldable origami is quite large, for example encompassing the
famous paper crane, similar to the model shown in
Fig.\ \ref{fig:twocolor}.

There are several theorems that have been proven about flat origami
\citep[\eg][]{Hull1994,Hull2002}.  The main flat-origami result that
we exploit in the present paper is the two-colorability of polygons
outlined by origami crease lines.  That is, two colors suffice to
color them so that no adjacent polygons share the same color.  Here
`adjacent' means sharing a crease line (not just a vertex).

To see why two colors suffice, consider the paper crane shown in
Fig.\ \ref{fig:twocolor}.  Both sides of it are shown, along with its
appearance when unfolded.  Each polygon is colored white or gray
according to whether the polygon is facing `up,' i.e.\ with the same
orientation as it did initially, or `down,' if it has been flipped
over.  This uniquely colors each polygon, and each crease does indeed
divide `up' from `down' polygons.

A work of flat origami can be thought of as a function (a continuous
piecewise isometry) mapping the unit square (the unfolded sheet at
right in Fig.\ \ref{fig:twocolor}) into the plane.  Each crease
produces a reflection, reversing the direction of the vector on the
paper perpendicular to the crease.  The function is defined on each
polygon by a sequence of these reflections.  The color in each polygon
corresponds to its parity, i.e. depending on whether the number of
reflections used to define the function on that polygon is odd or
even.  It can also be measured locally with the determinant of the
matrix defining the function on the polygon; we will also use this
latter definition in the cosmological case below.

Besides two-colorability, there are other properties that
flat-foldable crease patterns have.  For example, Maekawa's theorem
states that in a flat-foldable crease pattern, the numbers of
`mountain' and `valley' creases around a vertex (a junction of
creases) differ by two.  (A mountain crease becomes folded to form an
upward-pointing ridge; a valley crease is folded in the opposite way.)
In paper origami, two-colorability can be shown from Maekawa's
theorem \citep{Hull1994}, so it may apply in some form to cosmological
origami as well.

Even for paper origami, a difficult problem is to test that an
arbitrary crease-pattern is physically flat-foldable without the paper
intersecting any folds; this is an NP-complete problem
\citep{BernHayes1996}.  There are further results that, for instance,
describe the angles around vertices, but they depend on the
non-stretchability of the origami sheet, making them inapplicable to
the cosmological case.

\begin{figure}
  \begin{center}
    \includegraphics[scale=0.3]{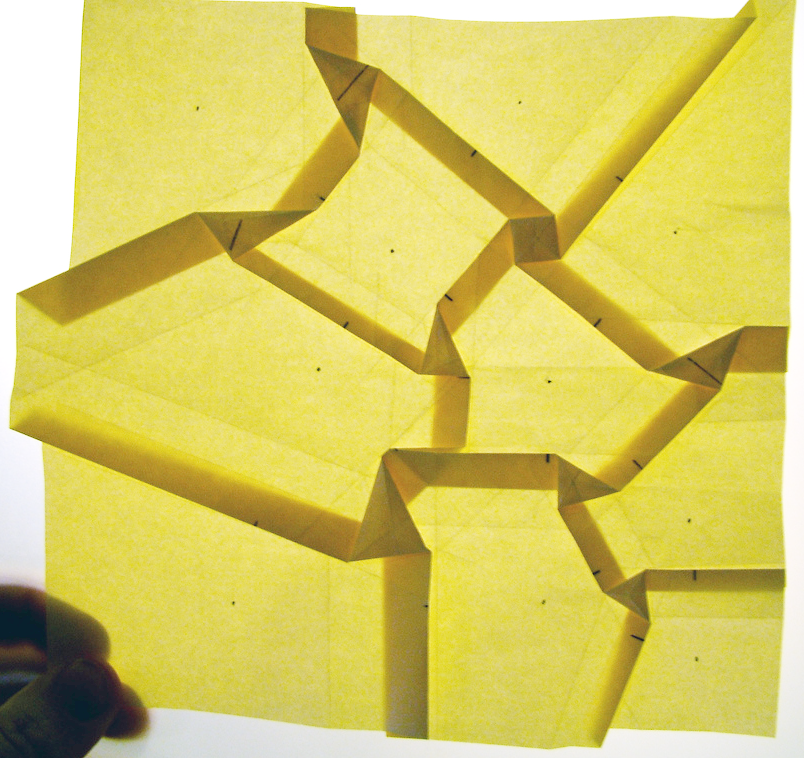}
  \end{center}  
  \caption{A Voronoi origami tessellation that bears some resemblance
    to cosmological voids, filaments and haloes.  If the work were
    unfolded, the polygons outlined by creases would be two-colorable
    according to which way the polygon is facing.  E.g.\ `voids' and
    the topmost polygons in `haloes' could be colored red, and the
    paper turned upside down within the `filaments' could be colored
    green.  The simple structure of `filaments' and `haloes' is as in
    the Zel'dovich approximation; the true universe has more
    complicated folding patterns, shown in subsequent figures.  Figure
    by Eric Gjerde (\url{http://www.origamitessellations.com/}), used
    with permission.
    \label{fig:vorigami}
  }
\end{figure}

Fig.\ \ref{fig:vorigami} shows a work of flat origami that bears some
resemblance to a network of filaments and clusters in cosmology.  Its
`voids' are Voronoi cells generated from the black points.  Voronoi
models of large-scale structure are good heuristic models of
cosmological structure formation
\citep[\eg][]{IckevandeWeygaert1987,KofmanEtal1990}.  The present
figure corresponds most closely to a Zel'dovich-approximation
evolution of particle displacements, in which structures fold up when
expanding voids collide, but overshoot and do not undergo realistic
further collapse.  Many origami tessellations are described, with
instructions, by \citet{Gjerde2008}.

\section{Application to Cosmological Structures}
\label{sec:oricosmi}
Moving from paper origami to cosmological structure formation
introduces a few changes.  The manifold (sheet) has three instead of
two dimensions.  It folds in six dimensions (three position, and three
velocity) instead of three.  The sheet also stretches inhomogeneously
in position space (all that we consider here), stretching more in
voids than in dense regions.  In phase space, on the other hand, there
is no stretching if using a symplectic metric, since by Liouville's
theorem, symplectic volumes are conserved.

A consequence of the position-space stretchiness of the cosmological
sheet is that, unlike in the paper-origami Fig.\ \ref{fig:vorigami},
creases need not extend indefinitely.  For example, a pancake may form
locally, without forming a full cosmic web.

With these caveats, folding of the three-dimensional sheet happens in
an analogous way to as in the two-dimensional sheet of paper
flat-origami.  Caustics within the three-dimensional cosmological
sheet are two-dimensional\footnote{As we discuss below, caustics that
  are lower-dimensional in Lagrangian space are in principle possible,
  but would require cylindrical or spherical collapse. In a realistic
  situation this never exactly happens; axes collapse one-at-a-time.},
just as caustics in two-dimensional origami paper are one-dimensional.

Importantly, the local parity on the sheet is easily defined, in the
same way as in flat origami.  At each position, define a first-order
approximation of the mapping of mass elements (particles) from initial
to final coordinates as a linear transformation (called the
deformation tensor) plus a translation.  If the spatial resolution in
Lagrangian space (i.e.\ the mass resolution in a simulation) is
sufficiently large compared to the scale on which Lagrangian space is
folding, this linear transformation is meaningful.  However, deep
within a halo, the spatial relationships between Lagrangian neighbours
in a practical $N$-body simulation may become essentially random, and
this condition may not be satisfied.

We may use the determinant $J(\bq)$ of this deformation tensor to
measure the parity \citep[\eg][]{whi09}; as mentioned above, this
definition can be used in two-dimensional flat origami, as well.  For
example, \citet{VogelsbergerWhite2011} use this parity measure to
explore the rococo fine-scale structure of caustics in Eulerian space.

$J(\bq)$ gives the volume of a mass element that gets deformed by the
tensor, times the parity.  The elements of the deformation tensor are
the components of initial-conditions Cartesian basis vectors
transformed to the final conditions,
\begin{equation}
  J(\bq) = \det \left(\delta_{ij}+\frac{\partial\Psi_i}{\partial q_j}\right).
  \label{eqn:deftens}
\end{equation}
Here $\bPsi$ is the displacement field.  For a mass element with
Lagrangian position $\bq$ and Eulerian position $\bx$, $\bPsi$ is
defined by $\bx(\bq) = \bq + \bPsi(\bq)$.  As usual in cosmology, we
use comoving coordinates.

\subsection{Streams and caustics in Lagrangian space}
Caustics and streams are usually considered in Eulerian space.  A
caustic is a fold in projected phase space where the density formally
goes infinite if particle discreteness is ignored, and the density is
smoothed on arbitrarily fine scales.  As for streams, at a given
Eulerian-space location $\bx$, there may be many of them; each stream
corresponds to a point in Lagrangian space that has ended up at $\bx$.

In Lagrangian space, by analogy to an unfolded origami sheet, we
define a stream as a contiguous three-dimensional region with the same
parity.  We define a caustic as a two-dimensional surface separating
streams from each other.  Defined this way, a caustic indeed
corresponds to a fold, since the parity swaps if one moves across it.

By definition, space is tessellated by streams that are outlined
by caustics.  The streams are also two-colorable, since the parity may
take only two values.

Two-colorability may seem hopelessly academic, and indeed it does not
have obvious observational consequences.  But in fact it greatly
restricts the arrangement of streams in three-dimensional Lagrangian
space.  For a generic arrangement of solids in three or higher
dimensions, the chromatic number (the number of colors required) is
bounded only by the number of solids.  To see this lack of bound,
consider a stack of arbitrarily long raw spaghetti.  Each may be
slightly rotated from the last, in a way that it touches all others
\citep{Guthrie1880}.  Now consider the connectivity of cooked
spaghetti; it is easy to imagine a high chromatic number, without any
special noodle-arrangement.

In summary, in two dimensions, origami-foldability reduces the
maximum-possible chromatic number from 4 to 2. In the
three-dimensional dark-matter sheet, foldability reduces the
maximum-possible chromatic number of an arrangement of solids far more
dramatically, from $\nstreams$ ($\sim10^{14}$ at the Sun's location
\citep{VogelsbergerWhite2011}) to 2.

\subsubsection{Graph-theory properties of the stream tessellation}
In graph theory, a two-colorable graph is called bipartite
\citep[\eg][]{ChromGraph2009}.  The vertices of our bipartite graph
are the three-dimensional stream regions, and the edges (linking
vertices together) are the caustic surfaces between them.  

At least one whole book is devoted to the subject of bipartite graphs
\citep{BGraphs1998}; here we list a few of their properties,
translating into cosmological terms.  First, there is no path
(stepping from stream to stream through caustics) starting and ending
at the same stream that consists of an odd number of steps.  Also, the
adjacency matrix of streams has several properties that arise from
bipartiteness.

Another result, K\"{o}nig's Minimax Theorem, pertains to matchings in
bipartite graphs.  In our case, a `matching' links pairs of streams
such that each stream has a unique `match' of the opposite parity.
The theorem states that the maximum possible number of matches equals
the `minimum vertex cover,' the minimum number of streams needed to
include in the graph such that all caustics touch at least one stream.

For example, imagine that only two isolated Zel'dovich pancakes form
in a universe.  In this case, there would be three Lagrangian streams:
one positive-parity stream consisting of everything except the
pancakes, and the two negative-parity collapsed patches.  There are
two caustics: the boundaries between the negative-parity islands and
the positive-parity sea around them.  The maximum matching of positive
to negative streams would have only one match: the initial-parity
stream with either of the pancakes.  By K\"{o}nig's Minimax Theorem,
this should equal the minimum vertex cover, and indeed it does: one
stream (the positive-orientation sea) touches both caustics.

Considering the dual graph, in which streams and caustics swap roles,
K\"{o}nig has another result.  His Coloring Theorem for bipartite
graphs applies to the dual graph of caustics joined by streams: the
chromatic number for the dual graph equals the maximum number of
caustics around a single stream.  If the graph of streams joined by
caustics were not bipartite, the dual graph would generally have a
larger chromatic number than the maximum number of caustics around a
single stream.  

\subsubsection{Discussion of the Lagrangian stream-caustic definition}
Before continuing on to measurements from simulations, we examine our
definitions of streams and caustics a bit further, and problems that
may arise from them.

A physical problem with our parity definition arises from a type of
folding that is, in principle, possible in cosmology, but not in paper
origami because the paper cannot stretch.  Cosmological caustics may
form in spherical or cylindrical collapse, not just planar collapse.
Like planar collapse, spherical collapse reverses parity, but
cylindrical collapse does not; it simply produces a 180$\degr$
rotation in the two axes perpendicular to the cylinder.

However, in a physically realistic situation, the probability that
more than one axis will collapse exactly simultaneously is zero.  In
the Zel'dovich approximation, for example, the deformation tensor will
never have two exactly equal eigenvalues.

In a practical $N$-body simulation, the finite temporal and mass
resolution will enable two- (and three-) axis crossings to happen
between timesteps.  In applying our parity definition to a simulation,
we unfortunately miss some such unresolved caustics. 

Another difficulty springs from numerical errors and errors
(e.g.\ two-body effects) due to particle discreteness in a practical
simulation.  After many dynamical times suffering deep within a halo,
Lagrangian mass elements may be hopelessly tangled for these reasons,
and not just because of the doubtlessly plentiful caustics that have
formed.  Indeed, as shown in the figures below, halo regions have
seemingly random parities, and it is not clear whether these parities
arise from physical caustics, or this numerical `noise.'

It may be difficult to visualize how a newly forming caustic forming
in an already high-density region (already with many overlapping
streams) behaves in Lagrangian space.  In Lagrangian space, the
caustic surface slices the various streams at various angles depending
on how each stream has been rotated, stretched and reflected.  But by
definition, the new streams are still three-dimensional regions,
divided by two-dimensional caustics.

\subsection{Simulation measurements}

\begin{figure*}
  \begin{minipage}{175mm}
    \begin{center}
      \includegraphics[scale=0.6]{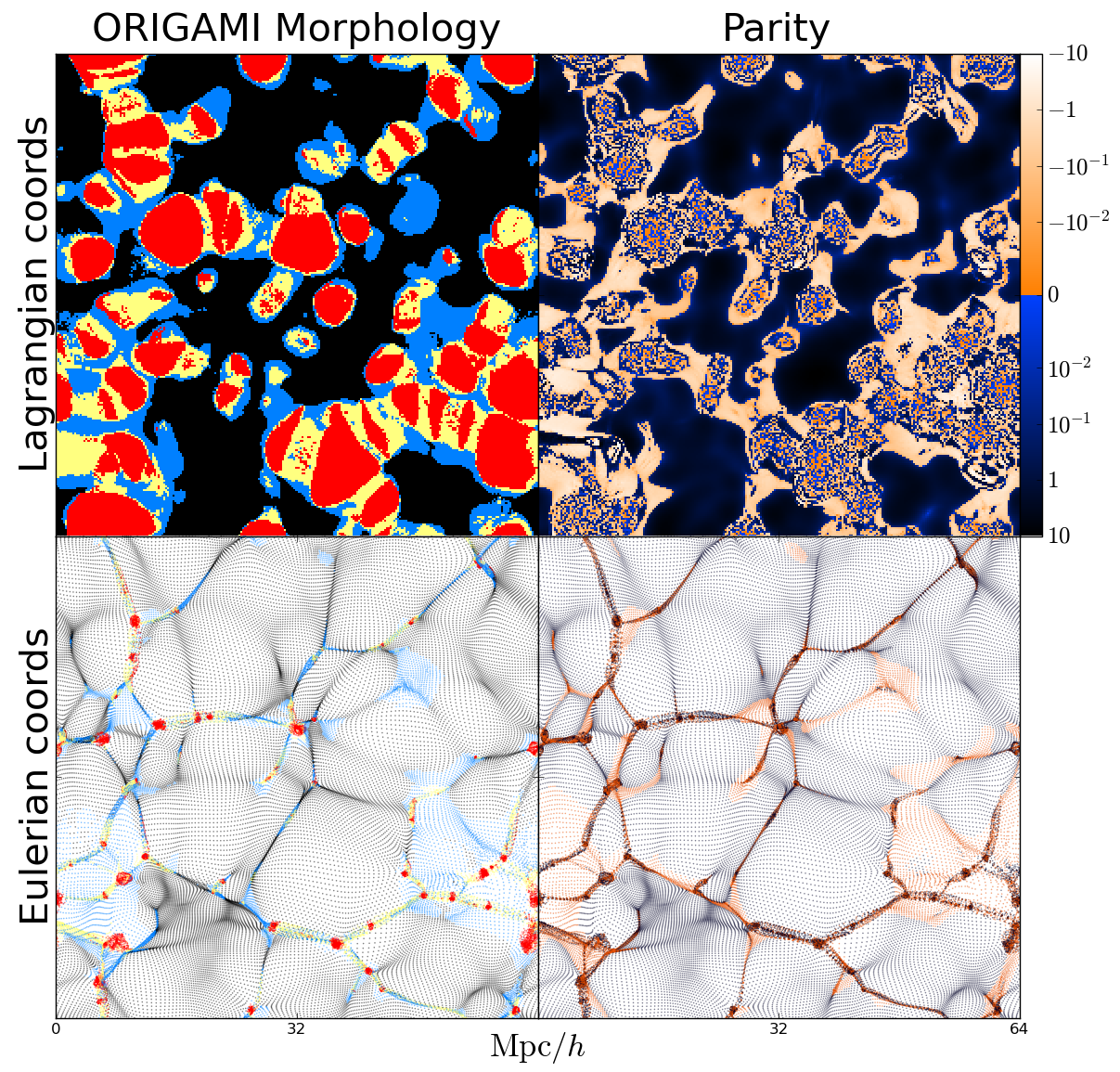}
    \end{center}  
    \caption{A 2D Lagrangian slice through a 3D Lagrangian
      cosmological sheet (top, unfolded; bottom, folded). Quantities
      were measured from a 256$^2$ sheet of particles from a
      256$^3$-particle \LCDM\ $N$-body simulation; The $256^2$
      particles share the same $z$-coordinate in the
      initial-conditions lattice, where $z$ points out of the
      page. Before running the simulation, the initial conditions were
      smoothed with a 1\hmpc\ Gaussian window, to inhibit small-scale
      structure formation.  Top panels use Lagrangian coordinates, in
      which each particle is a square pixel in a 256$^2$-pixel image.
      In the bottom panels, particles are shown in their actual
      present-epoch Eulerian $(x,y)$ coordinates, projecting out the
      $z$ coordinate (in which the slice does have some extent).  At
      left, void, wall, filament and halo \org\ morphologies are shown
      in black, blue, yellow and red, respectively.  In right panels,
      particles are colored according to $J$, i.e.\ the volume of
      their fluid element times its parity.  Black/blue particles have
      right-handed parity (as in the initial conditions), and
      white/orange particles have swapped, left-handed parity.  The
      color scale was stretched around zero with the function
      $x\to\sinh^{-1}(10^3x)$.}
    \label{fig:morphparity1}
  \end{minipage}
\end{figure*}

Fig.\ \ref{fig:morphparity1} shows the folding up of a two-dimensional
Lagrangian slice of $256^2$ particles from a three-dimensional
$256^3$-particle gravitational simulation run to the present
epoch. Pixels (in Lagrangian coordinates) and particles (in Eulerian
coordinates) are colored according to their \org\ morphologies (left)
and parities (right).

Panels at right show a measurement from an $N$-body simulation of the
quantity we use to determine the local parity, $J(\bq)$ from
Eq.\ \ref{eqn:deftens}.  This gives the volume of the mass element
(inversely proportional to its density) times its parity.  The parity
can be read off from the color scale.  Regions with right-handed
parity (as in initial conditions) are black or blue; left-handed
regions are white or orange.  Each pixel represents a particle. In
origami terms, this is the sheet before folding.  Note that the
magnitude is quite small in the cores of halo regions, because mass
elements shrink considerably in high-density halo regions.  To
estimate the tensor at each particle, we use the final-conditions
separations between the six particles that initially surround the
particle, i.e.\ $\left\{\bx_{i+1,j,k}-\bx_{i-1,j,k},
\bx_{i,j+1,k}-\bx_{i,j-1,k}, \bx_{i,j,k+1}-\bx_{i,j,k-1}\right\}$,
where the indices refer to positions on the Lagrangian particle grid.
Note that the resolution of this measurement is twice the smallest
available Lagrangian length scale (the interparticle spacing).

The morphology from the \org\ algorithm (not to be confused with the
origami-folding analogy itself) is shown in panels at left.  A
particle's \org\ morphology depends on the positions of many other
particles, not just (as in the parity measurement) its immediate
Lagrangian neighbours.  \org\ measures morphologies by comparing
particle orderings of along axes (rows and columns) in the initial
Lagrangian lattice to their orderings in Eulerian space.  Halo,
filament, wall and void particles have been crossed (compared to the
initial conditions) by some other particle along 3, 2, 1, and 0
orthogonal axes, respectively.  That is, if we index particles in a
row of the initial Lagrangian lattice, particles with Lagrangian
indices $i$ and $j$ have crossed if $i<j$ but their order along that
axis is swapped, i.e.\ their Eulerian positions $x_i>x_j$.  Note that
the total number of stream-crossings that a particle has undergone may
be arbitrarily high, but the number of perpendicular axes along which
these have happened is at most 3.

These \org\ morphologies, and the parity colorings, correspond quite
well to the locations of structures after folding.  There are a few
regions that would look voidy if they weren't colored, but are
colored nonetheless (blue at left, and orange at right).  These
particles have crossed others in the direction perpendicular to the
page.

The simulation shown was run to the present epoch using the {\scshape
  gadget-2} code \citep{Gadget2005} with 256$^3$ particles in a box of
size 64\hmpc, assuming a concordance \LCDM\ set of cosmological
parameters ($\Omega_b=0.04$, $\Omega_m=0.3$, $\Omega_\Lambda=0.7$,
$h=0.73$, $n_s=0.93$, $\sigma_8=0.81$).  Before running the
simulation, its initial-conditions density field was smoothed with a
Gaussian of width $\sigma=1$\hmpc, inhibiting small-scale structure
formation and clarifying the parity measurement.  This smoothing
reduced $\sigma_8$ (the dispersion in the overdensity within spheres
of radius 8\hmpc) by 0.4 per cent.  This small-scale power attenuation
would happen in the real universe if the dark matter had substantial
warmth; however, warm dark matter would also cause the
three-dimensional `sheet' it occupies to be smeared out by thermal
motions.

There is some agreement between outer caustics identified by
\org\ morphology (the boundaries between black and non-black regions)
and as defined by parity (the outermost boundaries between black/blue
and white/orange regions), but the agreement is not perfect.  There
are a couple of reasons for this.  First, the parity is defined on a
larger grid spacing on the Lagrangian lattice (two interparticle
spacings, not one).  Thus, if only one particle crossed another, this
would be detected by the \org\ criterion but not necessarily by the
parity criterion.  Second, the \org\ algorithm detects crossings
projecting along the Eulerian $(x,y,z)$ axes; in contrast, the parity
measurement uses only the local vectors between Lagrangian neighbours.
So, unlike the parity measurement, the \org\ morphology is in
principle sensitive to large-scale rotations (fortunately, small on
cosmological scales).  Still, the \org\ halo, filament and wall-finder
shines here, where the smoothed initial conditions produce visually
obvious morphologies.

Also note that streams need not have simple topologies.  For example,
the largest stream in Fig.\ \ref{fig:morphparity1} is just the `void'
stream, where no folding has occurs.  This stream may in principle be
entirely connected, and not split into isolated cells by caustics.
Regions where folding has occurred may be isolated `holes,' giving the
void stream a non-trivial topology.  Under our definition, curiously,
a newly formed pancake has just one caustic associated with it: the
entire surface separating the negative-parity region from the unfolded
stream, even if kinematically it might seem more natural to define two
caustics (the top and bottom of the pancake).

\begin{figure*}
  \begin{minipage}{175mm}
    \begin{center}
      \includegraphics[scale=0.6]{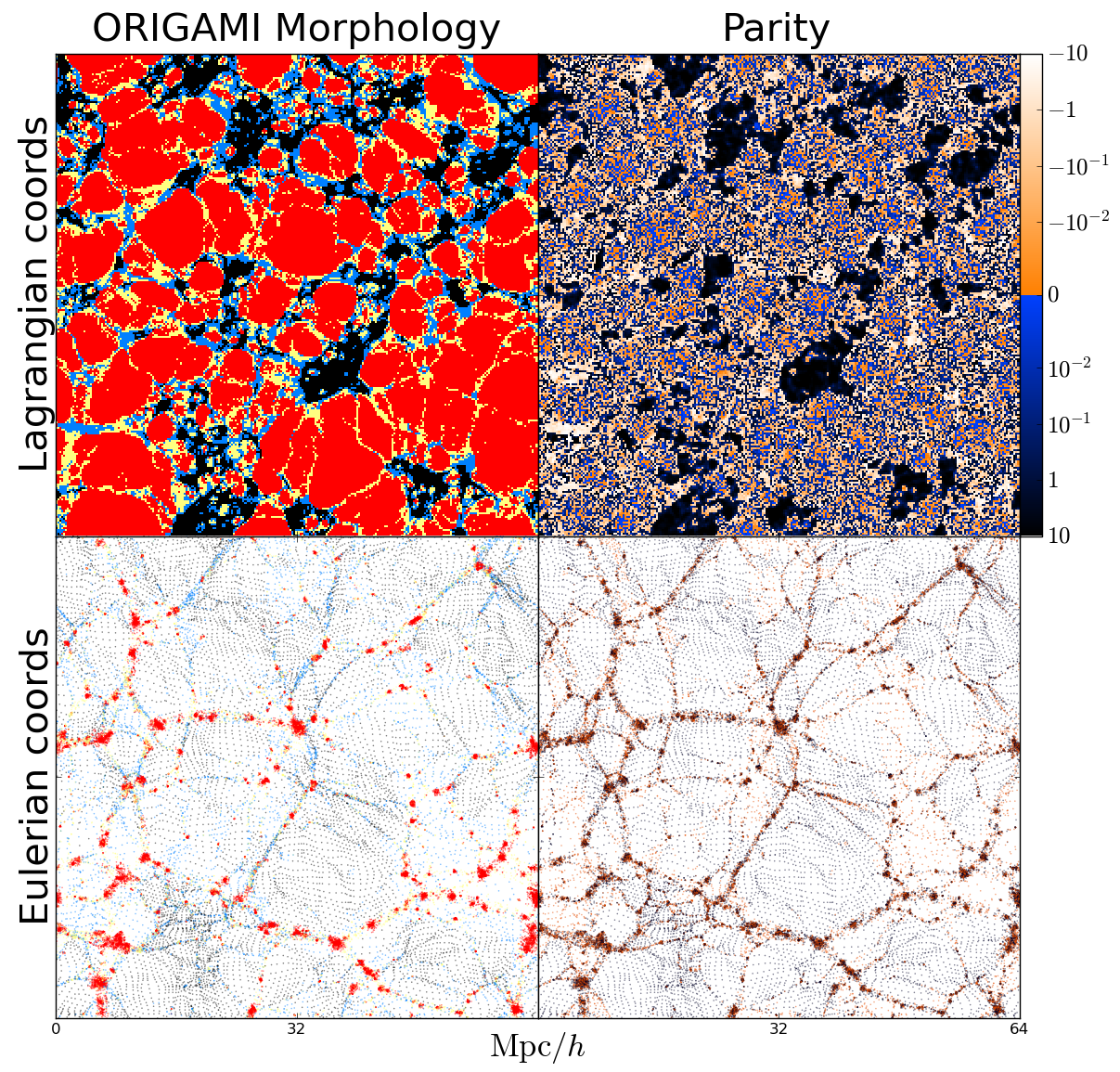}
    \end{center}  
    \caption[1]{ Same as Fig.\ \ref{fig:morphparity1}, except measured
      from a simulation with full initial power, i.e. unsmoothed
      initial conditions.}
    \label{fig:morphparity0}
  \end{minipage}
\end{figure*}

Fig.\ \ref{fig:morphparity0} is the same figure as
Fig.\ \ref{fig:morphparity1}, except that the simulation was run
without smoothing the initial conditions.  The small-scale structure
in Fig.\ \ref{fig:morphparity0} makes the parity map (the upper-right
panel) much more cluttered from additional structure.  There are many
visible extended streams (patches of identical parity), especially
voids, but much of the plot looks essentially random.  Halo particles
(comprising most of the particles) have likely undergone many shell
crossings.  We again caution, however, that some of this apparent
randomness could be from numerical `noise.'

\begin{figure*}
  \begin{minipage}{175mm}
    \begin{center}
      \includegraphics[scale=0.6]{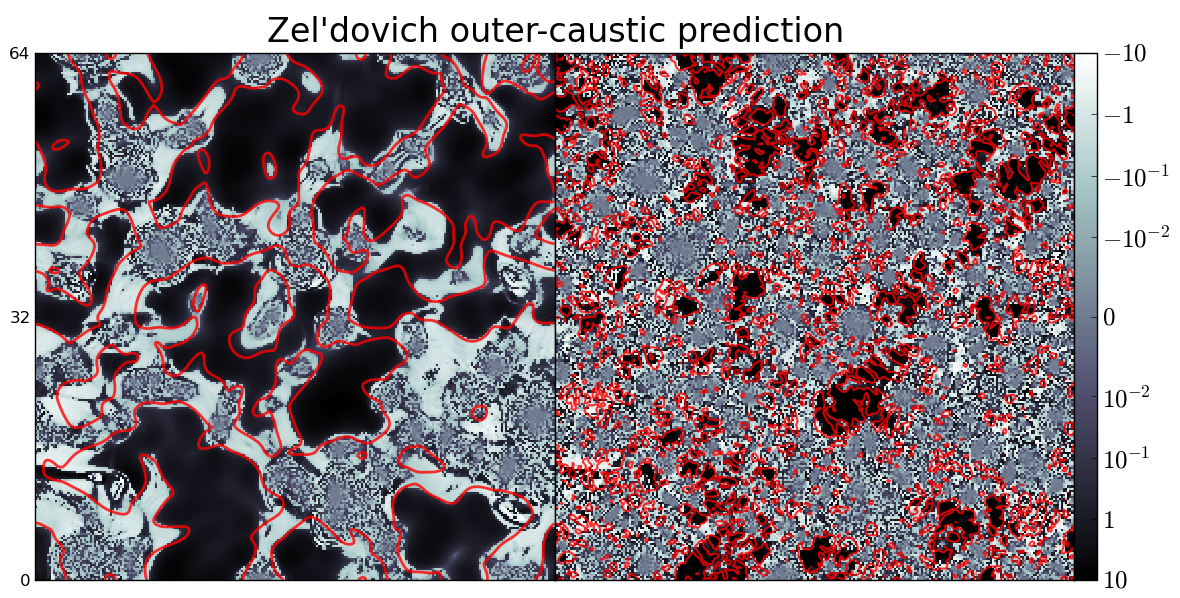}
    \end{center}  
    \caption{Comparison of a Zel'dovich prediction of where outer
      caustics should form (red contours) to where they actually do
      (boundaries between black and white regions), in the
      64-\hmpc\ simulations.  In the left panel, the initial
      conditions have been smoothed, but not in the right panel.
      Pixels (corresponding to particles) are colored according to
      the present-epoch deformation tensor determinant $J(\bq)$.  The
      plots under the red contours are same as the upper-right panels
      of Figs.\ \ref{fig:morphparity1} and \ref{fig:morphparity0},
      with a different color scale that hardly distinguishes small
      negative from small positive $J$.  The red contours show the
      Zel'dovich prediction, i.e.\ the zero-crossing of the linearly
      extrapolated largest eigenvalue of the initial deformation
      tensor.}
    \label{fig:zeldpred}
  \end{minipage}
\end{figure*}

Fig.\ \ref{fig:zeldpred} shows Lagrangian $J(\bq)$ plots, adding red
contours at the Zel'dovich-approximation \citep{Zeldovich1970}
prediction for the position of the outermost shell crossing.  In the
Zel'dovich approximation, the evolution with time $\tau$ of the volume
$V$ (in units of the mean) of a mass element at Lagrangian coordinate
$\bq$ is given by
\begin{equation}
  V(\bq,\tau) = \prod_{i=0}^2 \left[1-\lambda_i(\bq)D(\tau)\right],
  \label{eqn:zeldovich}
\end{equation}
a product over the three eigenvalues $\lambda_i$ of the deformation
tensor, using the linear growth factor $D(\tau)$.

For the Zel'dovich prediction in Fig.\ \ref{fig:zeldpred}, we
computed the largest eigenvalue of the deformation tensor (estimated
as above) at each particle in the initial conditions at redshift
$z=75$.  The red curve is the zero-contour of $[1-\lambda_{\rm
    max}D(z=0)]$, evolved forward to the present epoch with the
ratio of growth factors $D(z=0)/D(z=75)$.

The agreement is rather good between the Zel'dovich prediction and
actual outer-caustic locations for the smoothed simulation (left).
For the unsmoothed simulation, the agreement is worse, but still not
disastrous.  It is not surprising that the Zel'dovich approximation
works best at finding large-scale caustics.

\section{Conclusion}
In this paper, we further elucidate the analogy between structure
formation in cosmology and origami.  We illustrate an insight from
origami mathematics that appears to be applicable to cosmology: the
two-colorability of three-dimensional regions (streams) outlined by
two-dimensional caustic surfaces in the initial conditions.  That is,
two colors suffice to color them such that adjacent regions do not
have the same color, in the same way that four colors suffice to color
any planar map.  In fact, this result is rather trivial, if streams
and caustics in Lagrangian space are defined as we do.  However, a
general arrangement of three-dimensional regions has no bound on the
required number of colors, so this is a significant restriction on how
dark matter can assemble itself into structures.  Two-colorable graphs
such as this have many properties that may prove useful in
understanding structure formation.

While much is known about the behavior and morphology of caustics and
streams in Eulerian space, their behavior in Lagrangian space is also
of interest.  To our knowledge, this paper contains the first explicit
illustration of the shapes of caustics and the streams in Lagrangian
space.  Many questions can be asked about them.  What constraints
exist on their topologies?  Do Lagrangian caustics meet in
one-dimensional curves (or zero-dimensional points), and what do these
one- and zero-dimensional loci mean physically?

Admittedly, two-colorability in itself is lacking in obvious
observational consequences, but it is closely tied to the process of
caustic formation, of great interest for prospective dark-matter
direct and indirect detection.  As cosmological simulations push to
smaller and smaller scales, departing farther and farther from
comfortable linear-regime physics, it is useful to know as much
mathematically as possible (including the present result) about the
structures that develop.  It would be interesting to explore whether
enforcing the properties of caustics and streams found here, or of the
three-dimensional sheet structure itself, would be useful to `clean'
numerical noise away from simulations.

The origami analogy also has substantial pedagogical, public-outreach
value.  Crease-patterns of a popular hands-on `fold your own
galaxy' activity, similar to Fig.\ \ref{fig:vorigami}, can be found at
\url{http://skysrv.pha.jhu.edu/~neyrinck/origalaxies.html}.

\section*{Acknowledgments}

I thank Miguel Arag\'{o}n-Calvo for the use of the simulations
presented here, and for many discussions; Robert Lang for an inspiring
colloquium about paper origami, and valuable discussions; Tom Hull for
permission to use Fig.\ \ref{fig:twocolor}; Eric Gjerde for permission
to use Fig.\ \ref{fig:vorigami}; Alex Szalay and Bridget Falck for
fruitful discussions; and Tom Bethell for many useless discussions.  I
also thank the referee for insightful suggestions.  I am grateful for
support through Alex Szalay from the Gordon and Betty Moore
Foundation.

\bibliographystyle{hapj}
\bibliography{refs}

\end{document}